\documentstyle[aas2pp4]{article}

 \newcommand{\kms}{{\rm\,km\,s$^{-1}$}}

\newcommand{\Mpc }{$h^{-1}$\thinspace Mpc } \newcommand{\vbi}{v$_{BI}$}
\newcommand{\sbi}{$\sigma_{BI}$}

\begin{document}

\title{Shape of the Galactic Orbits in Clusters}

\author{Amelia C. Ram\'{\i}rez  \& Ronaldo E. de Souza } 
\affil{Astronomy Department, University of S\~ao Paulo, C.Postal 9638, SP 
01065-970, Brazil\\
e-mail aramirez@kbrita.iagusp.usp.br and ronaldo@kbrita.iagusp.usp.br \\
ftp:  cardeal.iagusp.usp.br anonymous /pub/astro/aramirez}
\slugcomment{Submitted to the Astrophysical Journal}


\begin{abstract}

A kinematical  analysis   applied  to a  sample of    galaxy  clusters
indicates that  the  differences between the velocity  distribution of
elliptical and spiral galaxies are  associated with the shape of their
orbit families.  The orbital anisotropies present
on each morphological population could be measured with the use of
a parameter  which is the ratio  of the radial and tangential velocity
dispersions, and can be recovered  through the observed  line-of-sight
velocity  distribution.    When a  Gaussian velocity   distribution is
assumed, having different dispersions along  the radial and tangential
directions,   we conclude that the    orbits of elliptical galaxies in
clusters are close to radial, while spirals have more circular shaped
or  isotropic orbits.    Lenticulars galaxies  shares  an intermediate
orbital parameter, between spirals and ellipticals. 

\end{abstract}


\section{Introduction}

As several works have suggested late type galaxies in clusters present
 a different kinematical behavior compared to early type objects.  The
 cluster velocity  dispersion   calculated  using only   spiral galaxy
 members seems to be always higher  than the one calculated using only
 elliptical galaxies, although at a low significance level.  Moreover,
 late type galaxies tend to present broader spatial distribution when
 compared to the one observed in  early type galaxies, which are more
 concentrated towards the cluster core.  According to Dressler (1980a)
 this effect is  due to  the   correlation  between  the frequency of
 morphological  types and the  local galaxy  density  of the clusters.
 However, Whitmore and Gilmore (1991) proposed that this morphological
 segregation effect depends upon radial distance to the cluster center
 instead of the local  density.   Another difference between late  and
 early  type galaxies in  clusters,  even if  marginally supported  by
 observations,  is  the mass segregation   reported  by Biviano et al.
 (1992).  They investigate the  velocity field of nearby clusters  and
 found  that    galaxies brighter than    the  third-ranked    object,
 preferentially bright elliptical  galaxies, tend  to present velocity
 dispersion lower than average.  They proposed that these objects have
 suffered the effect of a persistent dynamical friction process. 

The explanation to the existence of morphological segregation is
still  a  matter of  discussion  between  the  {\it nurture} and  {\it
nature}   scenarios of galaxy  formation  (Bower  1995).   Most common
interpretations are attributed to the possibility
that  spirals have been accreted  by the cluster  more recently, after
the    collapse and violent  relaxation  of  the initial population of
galaxies which now constitute the cluster core (Moss and Dickens 1977,
Tully  \& Shaya 1984, Huchra  1985, Dressler \& Shectman 1988, Sodr\'e
et al.  1989, Bird et  al.  1994, Andreon  1994, Biviano et al.  1997,
Colless   \& Dunn 1996, Fadda et   al.  1996,  Girardi   et al.  1996,
Scodeggio et al.  1995, Andreon   \& Davoust 1997 and Andreon   1996).
When a nurture scenario is assumed it  can be expected that during the
cluster life different kinematical  distribution of their members will
produce  different  interactions  and  probably different response  to
them.  Thus, the stability of the morphological shapes of galaxies, as
they plunge towards  the central regions, will  depend mostly on their
orbits.   Some  objects, with roughly  circular  orbits, will not have
their morphologies  seriously affected because  they avoid the cluster
center  where the probability  of occurring an strong interaction will
be larger.   However, those with more  eccentric orbits will cross the
densest  cluster  regions and will  experience  on average an stronger
environmental influence and a higher encounter rate. 

In  this   paper we  present  a  kinematical  analysis of   spiral and
elliptical galaxies in  nearby  rich clusters, showing evidences  that
they  represent populations with   different families of  orbits.  Our
results impose a further restriction on plausible models of cluster of
galaxies formation, where each  morphological class must reproduce the
corresponding velocity field anisotropy. 

In section 2 is presented a discussion of the distribution function we
used  to analyze the line-of-sight  velocities  as a  function of  the
adopted anisotropy parameter.  In section 3 we define
 moment statistics over the velocity distribution. 
   In  section   4, we  use  the average     deviation of the
line-of-sight velocity |  the average  of  the absolute  value of  the
deviation  from the mean normalized  to the velocity dispersion of the
cluster  |  to trace the  orbit   distributions of  the elliptical and
spiral  galaxies of a sample of nearby galaxy
clusters.   In section  5  the  velocity dispersion  profiles  are  
 used to
reinforce that our conclusions are in agreement with  the best fits to
observed  clusters   made by  other authors.    In the   section  6 we
discussed some implications  of elliptical galaxies in clusters having
more eccentrics    orbits.   Finally section  7   summarizes  our main
conclusions. 

\section{Distribution of velocities on a isotropic Gaussian field}

Theoretical models   of  rich  cluster  formation   predict virialized
systems with Maxwellian  or nearly Maxwellian velocity  distributions,
resulting therefore in distributions  for the line-of-sight velocities
closely described by a Gaussian (Saslaw  1990, Ueda et  al. 1993).  If
clusters  are formed   by   mergers of   sub-clumps units,    computer
simulations  suggest  that tidal  interactions   may quickly drive the
total gravitational potential toward isothermality. On the other hand,
clusters with  an isothermal dark matter halo  should be single peaked
in number density having   spherical or elliptical  symmetric  shapes,
with  no  correlation between  the  position  and  velocity of  member
galaxies, having   therefore  velocity dispersion  independent  of the
radial position (Roettinger et al.  1993, Katz \& White, 1993). 

The above mentioned  models   leads to    the simplest description  of
clusters  structures. However  many clusters,  having in  some cases a
large number of measured velocities, present  a non-Gaussian projected
 velocity
distribution (Zabludoff et al.  1993).  These deviations may imply  in
anisotropic galaxy orbits and/or mixing of  two or more sub-population
of galaxies (Merritt 1988, Bird 1994, Merritt  \& Gebhardt, 1995).  In
this context, the  overall distribution could be  non Gaussian even in
the presence  of different populations,  represented by early and late
type galaxy members, having    individually a Gaussian   distribution.
Furthermore, different shape of the orbits could produce an additional
deviation from gaussianity.  To test  these hypothesis, we studied the
velocity distribution of  each morphological  population to verify  if
they show signs of non-isotropic distributions. 

To  study   the velocity distribution    of  a population,  being less
 restrictive as possible, Merritt \& Gebhardt (1995) used directly the
 observed data, fitting  their velocity distribution profiles with few
 or any parameter at all in order to obtain a unique solution.  In our
 case however, since  we  are interested in  the study  of the orbital
 anisotropy we rather prefer to use a parameterized statistics related
 to    a Gaussian  distribution.   Even   if  that assumption  may not
 accurately  represent the  true   velocity  distribution due  to   its
 simplicity, it can  be useful as  a first approximation to understand
 the effects of the anisotropy in the overall cluster structure. 

We    assume that   for  a  given   morphological   class the velocity
distribution   function  is     Gaussian,   having however   different
dispersions  along the  radial ($\sigma_R$)  and transversal directions
($\sigma_\perp$).  The  behavior of the  velocity distribution of this
system can therefore be described by the  anisotropy parameter $\eta =
\sigma_R/\sigma_\perp$.     A large value   of   $\eta$  for  a  given
population  means that its members  are  crossing the cluster with  an
almost radial  orbit,   and therefore are    more sensitive to  suffer
gravitational  encounters with objects  in  the dense central regions.
On the contrary, objects with lower  anisotropy parameter tend to have
a more circular orbit with small penetration in the dense core region.
Therefore,  on a given morphological   class the anisotropy  parameter
should allow us     to connect the    efficiency of  the   environment
perturbations with the related kinematical orbital behavior. 

At the  present it is  not  yet well understood   the influence of the
projection effects, due to foreground and background galaxies, and the
cluster substructure on  the studies of  cluster dynamics. However  we
could expect that their presence should  introduce some deviation from
the simplest  cluster models  scenario.   We  will consider  the
dynamical  behavior of a  cluster  without  taking into account  these
effects, but we return to this point late in the section 4. 

For  a radially symmetric cluster  the velocity distribution along the
line-of-sight   can be  derived   from  the assumed  spatial  velocity
distribution.   Then, for a given morphological  class we will further
assume  that the anisotropy $\eta$  parameter have  a fixed value, and
consequently the probability of finding  an object at a given position
and velocity is determined by the expression, 

\[ dN(R,\theta,\phi;v_R,v_\theta,v_\phi)= n(R) R^2 \sin \theta dR
d\theta d\phi \times \]

\[ \frac{1}{(2\pi)^{3/2} \sigma_R \sigma_\perp^2}
e^{-\frac{v_R^2}{2\sigma_R^2} - \frac{(v_\phi^2 +
v_\theta^2)}{2\sigma_\perp^2}} dv_R dv_\phi dv_\theta \]

\noindent  where  $n(R)$ is the density  profile.   Since the velocity
distribution does not  depend on $R$, and  the density is  spherically
symmetric, we  can easily integrate  the radial  contribution and also
the terms related to the spatial azimuthal coordinate to obtain, 

\[ dN(\theta;v_R,v_\theta,v_\phi)= \frac{N}{2}\sin \theta d\theta
\frac{1}{(2\pi)^{3/2} \sigma_R \sigma_\perp^2} \times \]

\[ e^{-\frac{v_R^2}{2\sigma_R^2} - \frac{(v_\phi^2 +
v_\theta^2)}{2\sigma_\perp^2}} dv_R dv_\phi dv_\theta \]

\noindent where $N$ represents the total number of objects in the
system.

In order to obtain the observed velocity  distribution a better choice
is to use the cylindrical coordinates $v_r,  v_z, v_\phi$, where $v_z$
is directed along the  line-of-sight.  The transversal component to the
line of sight, $v_\phi$, is independent of the angle $\theta$, and can
be integrated to obtain, 

\[ dN(\theta;v_r,v_z)= \frac{N}{2} \frac{1}{2\pi \sigma_R \sigma_\perp
} e^{-\frac{v_R^2}{2\sigma_R^2} - \frac{v_\theta^2}{2\sigma_\perp^2}}
\sin \theta d\theta dv_r dv_z \]

This  equation  can be  further integrated  using the transformations $v_R=v_r
\sin \theta +v_z  \cos \theta$, and $v_\theta  = v_r \cos \theta - v_z
\sin   \theta$,   and  changing  variables   to $p=v_r/\sigma_R$   and
$q=v_z/\sigma_R$.  Using these  relations we can express  the observed
velocity distribution along the line-of-sight in the form, 

\[ \frac{d N(q)}{dq} = \frac{N}{2} \int^\pi_0 \sin \theta
\int_{-\infty}^\infty \frac{\eta}{2\pi} \times \]

\[ e^{-\frac{(p \sin \theta + q
\cos \theta)^2}{2} - \frac{\eta^2 (p \cos \theta - q \sin
\theta)^2}{2}} dp d \theta \]

After  some manipulation to complete  the square  term entering in the
internal expression, the integral  related to the velocity  term ($p$)
can be solved  exactly.  Since the  mean expected velocity  dispersion
can be    expressed  in  the  form    $\sigma  = \sqrt{(\sigma_R^2   +
2\sigma_\perp^2)/3}$, we   may use  it  to  define a  more interesting
reduced variable  $u = v_z/\sigma$.   Introducing the symbol $\omega =
\cos \theta$,   we finally obtain   the line-of-sight velocity density
distribution in the form, 

\begin{equation} F(u;\eta ) = \frac{1}{(2\pi)^{1/2}}
			  \int^1_0 \frac{1}{\Theta(\omega,\eta)}
			 e^{-\frac{u^2}{2\Theta(\omega,\eta)^2}}
			 d\omega
\end{equation}

\noindent         where      the       term      $\Theta(\omega,\eta)=
\sqrt{\frac{3(1-\omega^2 +\eta^2\omega^2)}{2+\eta^2}}$  represents   a
correction of  the  velocity dispersion  due to   the presence  of the
anisotropic field.  This  equation  corresponds to the   line-of-sight
velocity distribution in a Gaussian velocity field with anisotropy and
can only   be  solved   numerically. In   Figure  1 we    present this
distribution for   some  representative   values  of the   anisotropic
parameter.   In  particular,  for $\eta=1$ we   retrieve  the expected
Gaussian shape for an isotropic velocity  field.  We remark that, even
if the radial and transversal distributions are assumed to be Gaussian,
the observed distribution along the line-of-sight is not Gaussian when
the anisotropy parameter is different of one. 

\begin{figure}[ht] 
\plotone{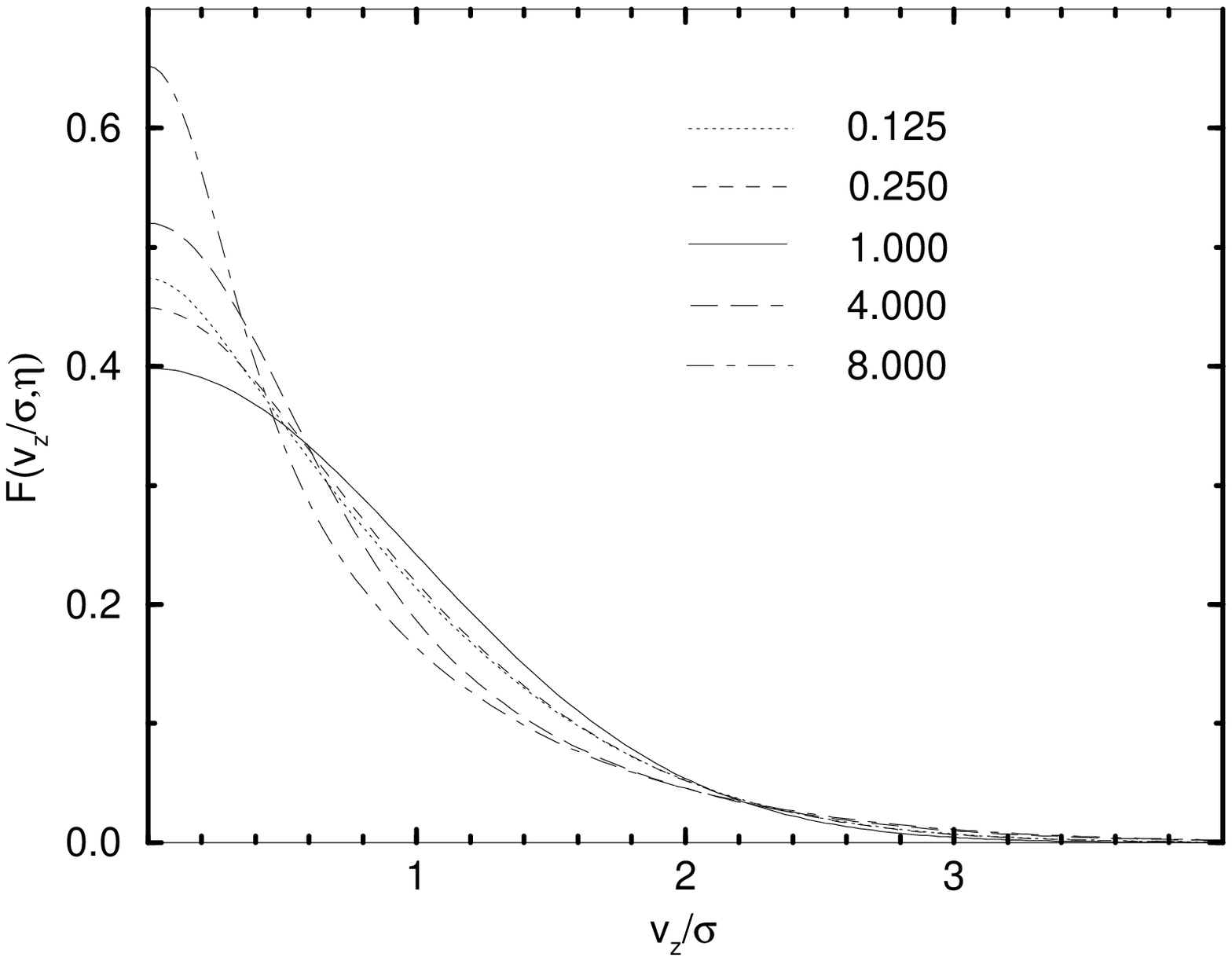} 
\caption
{Velocity distribution along the
line-of-sight normalized to the velocity dispersion for different
values of the anisotropic parameter. The smallest central peak value, and 
also the
kurtosis,  corresponds  to the case of an isotropic distribution
(continuous line), that results in a Gaussian curve. \label{f1}} 
\end{figure}

\placefigure{f1}

\subsection{Average deviation, kurtosis and peak value of $F(u;\eta)$}

Although in   the general  case  the   line-of-sight velocity  density
distribution can   be  estimated only   by  numerical  methods,  it is
interesting to note that the expressions for the moments can be solved
exactly.  The distribution $F(u;\eta)$  is symmetric by  construction,
resulting  that  the first  centered  moment  is  obviously zero.  The
second moment corresponds to the  variance of the distribution and can
be estimated by the expression, 

\begin{equation} \sigma^2_u = \int_{-\infty}^\infty u^2
\frac{1}{(2\pi)^{1/2}}
			  \int^1_0 \frac{1}{\Theta(\omega,\eta)}
			 e^{-\frac{u^2}{2\Theta(\omega,\eta)^2}}
			 d\omega du
\end{equation}

\noindent inverting the order of  integration we can easily show  that
the   function  $\Theta$ is    eliminated    before doing the   second
integration,   resulting    that  the  velocity   dispersion   remains
independent of the  anisotropic parameter.   Therefore, we may  expect
that   two populations responding  to the  same  potential can present
different orbital  shapes,   but  their velocity dispersions   remains
constant. 

The  central peak  value of the  distribution $F(0;\eta)$  may also be
easily integrated resulting in the expression, 

\begin{equation} 
F(0;\eta) = \sqrt{\frac{2+\eta^2}{6\pi(\eta^2 -1)}} \ln(\sqrt{\eta^2-1}+\eta) \;\;\;\;\;\;\;\;\; \eta^2 >1 
\end{equation}

\begin{equation} 
F(0;\eta)=\sqrt{\frac{2+\eta^2}{6\pi(1-\eta^2)}} sin^{-1}(\sqrt{1-\eta^2}) \;\;\;\;\;\;\;\;\; \eta^2 <1 
\end{equation}

An interesting feature is that the central  value of this distribution
is bounded to a limiting  value $\frac{1}{\sqrt{6\pi}}\ln (2\eta)$ for
$\eta \gg 1$, while for $\eta \ll 1$ the central density approaches to
the limit $\sqrt{\pi/12}\simeq 0.51$.    For  $\eta = 1$  the  central
value tends to   $1/\sqrt{2\pi   }$, characteristic of    the Gaussian
distribution. 

We  may also   obtain the  fourth  order moment  and  consequently the
kurtosis, 

\begin{equation} 
K = \frac{12}{5}\left( \frac{\eta^2 - 1}{\eta^2 + 2}\right)^2 
\end{equation}

For highly radial  orbits ($\eta \gg 1$)  the kurtosis  approaches the
limit $K  \simeq 2.4$,  while  for circular orbits  ($\eta  \ll 1$) $K
\simeq 0.6$.  Moreover the minimum kurtosis ($K =  0$) is attained for
$\eta =1$ when the distribution is normal. 

A disadvantage     of using the    kurtosis, or  the   $F(0;\eta)$, as
indicative for   the  anisotropy parameter,  is  their  large sampling
error.  To overcome this difficulty we propose the  use of the average
or mean deviation (Kendall, Stuart and Ord, 1987) of the line-of sight
velocity   normalized   to   the   velocity    dispersion  ($|u|     =
<\frac{|v-\bar{v}|}{\sigma}>$).     The   predicted  value    of  this
statistical parameter can be estimated by the expressions, 

\[ |u| = \sqrt{\frac{6}{\pi (\eta^2+2)}} \biggl( \frac{\eta}{2} + \frac{1}{2} \frac{1}{ \sqrt{1-\eta^2}} sin^{-1}(\sqrt{1-\eta^2} ) \biggr)  \]

$\eta < 1$

\[ |u|  = \sqrt{\frac{6}{\pi (\eta^2+2)}} \biggl( \frac{\eta}{2} + \frac{1}{2} \frac{1}{ \sqrt{\eta^2 - 1}} ln(\sqrt{\eta^2 - 1} + \eta ) \biggr)  \]

$\eta > 1$

For purely radial orbits ($\eta  \gg 1$) $|u| \rightarrow \sqrt{3/2\pi
} \simeq 0.69$, while   for  circular orbits  ($\eta   = 0 $)  $|u|  =
\sqrt{3\pi  }/4 \simeq 0.77$.  In  the  case of $\eta   = 1$ we have a
Gaussian  distribution and we recover  the value  $|u| = \sqrt{2/\pi }
\simeq 0.80$. 

\begin{figure}[ht] 
\plotone{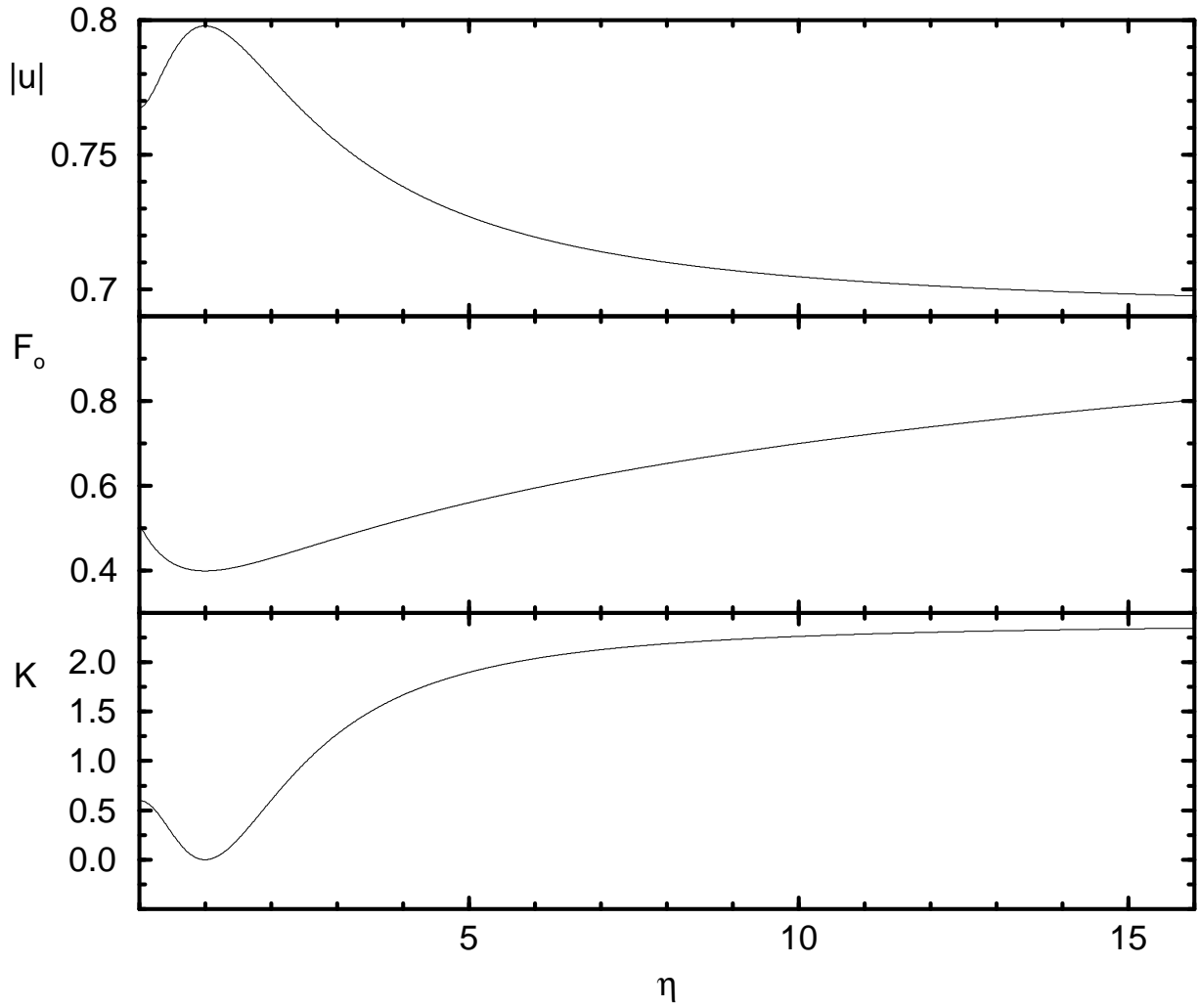} 
\caption
{Behavior of kurtosis, central peak value and mean velocity modulus for a
family of distribution functions, having the same velocity dispersion but
different values of the anisotropy parameter values.  \label{f2}}
\end{figure}

\placefigure{f2} 

In Figure 2 we  show these three statistical  parameters as a function
of   $\eta$.  We observe   that they   have  an extrema   at $\eta=1$,
producing  an indetermination  when we try   to derive $\eta$ from the
estimated values of these quantities.  In fact, using these parameters
we cannot distinguish between a purely circular model ($\eta = 0$) and
another one  having a  radial  contribution slightly  higher  than the
isotropic case.  The worse indetermination occurs when we use the peak
value statistics,  since in that case  we cannot distinguish between a
circular  orbit model and  another  one  having  $\eta  \leq  4$.  The
kurtosis  and the  average deviation   of the line-of-sight  velocity,
reduce this indetermination to  a region with $\eta  \leq 2$.  Due  to
this  limitation we will restrict our  analysis to a discussion of the
kurtosis  and the  average deviation.   However,  it is interesting to
note that highly radial models can  be easily discriminated, using any
one of these indicators, since in that  case the velocity distribution
is highly peaked, as can be seeing also from Figure 1. 

\section{The sample of galaxy clusters}

A sample of nearby rich clusters was selected to test for the presence
of  systematic   differences    of   orbital   parameters  among   the
morphological populations. We have collected material for all clusters
with $z   < 0.055$ from a  catalogue  of  measured redshifts  of Abell
clusters   compiled  by Andernach  (1991).    This  redshift limit was
adopted   in    order    to  avoid    dealing    with    morphological
misclassification.  From  a total of 1059  clusters only 323  have $z<
0.055$, most  of  them with less  than 10  velocities published in the
literature. We selected  for our tests only  those having  at least 65
objects with measured velocities   and morphologically classified   as
elliptical, spiral  or lenticular galaxies, within  2.5 \Mpc  from the
cluster center.    We   have  also discarded clusters    with  obvious
substructures, since in  those  cases the velocity distribution  could
result from a complex association of several small groups. Half of the
clusters  with at least 65  members with radial  velocities don't fill
the morphological  requirement which  is  essential for  our purposes.
Therefore,   we reach a  final  sample  having  only 18 clusters  that
satisfied the velocity, spatial and population requirements. 

The association of a galaxy with a given  cluster was decided on basis
of the  following procedure.  For each  cluster we have  estimated the
mean heliocentric velocity, and the  raw velocity dispersion along the
line-of-sight.  This was determined taking  into account all  galaxies
within 5.0 \Mpc of the nominal  center of each cluster and considering
objects of all morphological classes.  The sample of velocity data for
each  cluster   was  selected using  the  HEASARC\footnote{High Energy
Astrophysics Science Archive  Research Center Online Service, provided
by the NASA/Goddard Space Flight Center} facilities and completed with
the  CfA  Redshift  Catalog   (ZCAT/version November/1995, Huchra   et
al.   1992).   The velocities published  by   Biviano  et al.  (1996),
Colless  \& Dunn, (1996), and the  PGC Catalog (Paturel  et al.  1989)
were also used.  Non-member  galaxies  were detected by the  classical
3-$\sigma$ test (Yahil \& Vidal, 1977).   Finally, we concentrated our
estimatives   to an  outer radius   of  2.5 \Mpc  ~and selected  only
elliptical, spiral and lenticular galaxies.  At this point, new values
of mean velocity and velocity dispersion for the  cluster and for each
morphological    population were obtained.   All  the  mean values and
dispersions (location  and  scale values   in the  robust  statistical
notation)  were calculated with  the bi-weighted estimators, using the
ROSTAT\footnote{Version obtained from the ST-ECF Astronomical Software
Library ftp://ecf.hq.eso.org/pub/swlib}
 program  which  contains the versions  of  statistical routines
tested by T. Beers, K. Flynn, and K. Gebhardt for robust estimation of
simple statistics and described in Beers et al. (1990). All the errors
bars appearing in  this paper are at  the  68\% confidence  level, and
were obtained via  a     bootstrap resampling  procedure   with   1000
iterations. 

 The morphological classification was extracted from Dressler (1980b),
and The Third  Reference    Catalog  of Bright Galaxies     (RC3)  (de
Vaucouleurs et al.  1991).
 
\begin{deluxetable}{lrrrr@{~ }rr@{~ }rr@{~ }rr@{~ }rr@{~ }rl} 
 
\small
\tablewidth{0pc}
\tablecaption{Cluster kinematic parameters using the elliptical, spiral and 
         lenticular galaxies}
\tablehead{
\colhead{Name}      &
\colhead{N}         &
\colhead{$r_{200}$} &
\colhead{$r_h$}     & 
\multicolumn{2}{c}{v$_{BI}$ }      &
\multicolumn{2}{c}{$\sigma _{BI}$ }& 
\multicolumn{2}{c}{K}              &
\multicolumn{2}{c}{$|u|_{cl}$}     &
\multicolumn{2}{c}{$\sigma_{|u|}$} &
\colhead{References} \\
\colhead{}    &
\colhead{}    &
\colhead{Mpc} &
\colhead{Mpc} &
\multicolumn{2}{c}{km/s}&  
\multicolumn{2}{c}{km/s}& 
\multicolumn{2}{c}{}    & 
\multicolumn{2}{c}{}    & 
\colhead{}                       
}
\startdata
{\bf All}   \nl
  A1656&  458&  1.8 & 0.9 &   6969 & 48&  1032&  39&   0.8& 0.22&     0.71& 0.03& 0.59  &  0.03   & zcat; CD95\tablenotemark{a} \&B96\tablenotemark{b} \nl 
  Virgo&  402&  1.4 & 1.4 &   1380 & 39&   785&  23&  -0.6& 0.24&     0.76& 0.03& 0.55  &  0.02   & zcat           \nl
  A3526&  287&  1.5 & 1.2 &   3436 & 51&   863&  34&  -0.7& 0.29&     0.75& 0.04& 0.60  &  0.02   & zcat; Centaurus\nl
  A0194&  146&  0.7 & 0.9 &   5338 & 33&   398&  28&   1.1& 0.41&     0.71& 0.05& 0.64  &  0.06   & zcat           \nl
  A0548&  128&  1.5 & 1.4 &  12407 & 77&   870&  45&  -0.9& 0.43&     0.83& 0.05& 0.52  &  0.03   & zcat           \nl
  A1060&   98&  1.1 & 0.7 &   3668 & 64&   631&  44&  -0.5& 0.49&     0.74& 0.06& 0.59  &  0.04   & zcat; Hydra    \nl
  A2151&   93&  1.3 & 1.3 &  11000 & 76&   735&  46&  -0.7& 0.51&     0.80& 0.06& 0.59  &  0.04   & zcat; Hercules \nl
  A1644&   89&  1.6 & 1.6 &  14129 & 99&   934&  84&   0.3& 0.52&     0.67& 0.06& 0.59  &  0.09   & zcat           \nl
  A0539&   83&  1.2 & 0.8 &   8721 & 79&   715&  72&   0.2& 0.54&     0.66& 0.07& 0.67  &  0.07   & zcat           \nl
  A463s&   79&  1.1 & 1.1 &  12275 & 71&   624&  37&  -0.9& 0.55&     0.83& 0.06& 0.51  &  0.04   & zcat; DC0428-53\nl
  A0496&   77&  1.1 & 0.6 &   9836 & 71&   620&  67&   0.4& 0.56&     0.64& 0.08& 0.66  &  0.09   & zcat           \nl
       &  138&      &     &   9870 & 62&   728&  49&   0.0& 0.42&     0.70& 0.05&       &         & M92\tablenotemark{c} \nl  
  A3376&   77&  1.3 & 1.1 &  13909 & 84&   737&  72&   0.3& 0.56&     0.70& 0.08& 0.65  &  0.07   & zcat; DC0559-40\nl
  A805s&   77&  0.7 & 0.7 &   4351 & 48&   419&  38&   0.1& 0.56&     0.71& 0.07& 0.57  &  0.06   & zcat; DC1842-63\nl
       &  114&      &     &   4513 & 47&   503&  44&   0.1& 0.46&     0.66& 0.06&       &         & M92\tablenotemark{c} \nl
  A0119&   76&  1.5 & 0.9 &  13324 & 97&   840&  91&   0.0& 0.56&     0.66& 0.08& 0.68  &  0.08   & F93\tablenotemark{d} \nl
  A0754&   72&  1.4 & 1.4 &  16257 & 96&   812&  84&  -0.1& 0.58&     0.70& 0.08& 0.64  &  0.08   & zcat \nl
  A1983&   72&  1.1 & 1.5 &  13562 & 78&   660& 153&   1.5& 0.58&     0.52& 0.07& 0.56  &  0.10   & zcat \nl
  A1631&   68&  1.2 & 1.5 &  13971 & 85&   696&  50&  -0.6& 0.59&     0.79& 0.07& 0.58  &  0.04   & zcat \nl
 Fornax&   66&  0.6 & 0.7 &   1483 & 41&   330&  24&  -0.6& 0.60&     0.76& 0.07& 0.58  &  0.05   & F89\tablenotemark{e}  \nl \nl
{\bf Spirals}  \nl  \nl
  A1656&  189&  1.9 & 1.0 &   7036 &   79&  1082&  66&   0.7&  0.36&     0.73& 0.05& 0.61  &  0.04   &  \nl
  Virgo&  265&  1.4 & 1.6 &   1414 &   51&   824&  26&  -0.8&  0.30&     0.82& 0.03& 0.57  &  0.02   &  \nl
  A3526&  155&  1.5 & 1.7 &   3390 &   67&   838&  51&   0.6&  0.39&     0.71& 0.05& 0.59  &  0.04   &  \nl
  A0194&   76&  0.7 & 1.4 &   5306 &   47&   410&  32&  -0.2&  0.56&     0.78& 0.06& 0.63  &  0.11   &  \nl
  A0548&   45&  1.7 & 1.4 &  12407 &  151&  1002&  81&  -1.1&  0.73&     0.97&  0.09& 0.59  &  0.05   &  \nl
  A1060&   36&  1.1 & 1.1 &   3506 &  103&   611&  50&  -0.6&  0.82&     0.72& 0.10& 0.57  &  0.08   &  \nl
  A2151&   45&  1.2 & 1.1 &  11265 &  106&   706&  64&  -0.4&  0.73&     0.72& 0.09& 0.65  &  0.05   &  \nl
  A1644&   23&  1.7 & 1.8 &  13808 &  211&   987& 171&  -0.1&  1.02&     0.68& 0.13& 0.69  &  0.11   &  \nl
  A0539&   41&  1.0 & 1.6 &   8726 &   94&   597&  85&   0.1&  0.77&     0.53& 0.08& 0.50  &  0.09   &  \nl
  A463s&   18&  1.3 & 1.7 &  12215 &  186&   754& 114&  -1.0&  1.15&     0.95& 0.17& 0.67  &  0.10   &  \nl
  A0496&   18&  1.1 & 1.2 &  10055 &  150&   612& 102&  -0.8&  1.15&     0.74& 0.13& 0.63  &  0.15   &  \nl
  A3376&   24&  1.3 & 1.3 &  13845 &  164&   779& 122&  -0.1&  1.00&     0.75& 0.13& 0.60  &  0.12   &  \nl
  A805s&   39&  0.8 & 1.7 &   4395 &   70&   434&  63&   0.0&  0.78&     0.70& 0.11& 0.65  &  0.10   &  \nl
  A0119&   15&  1.7 & 1.3 &  13687 &  261&   959& 143&  -1.0&  1.26&     0.94& 0.17& 0.68  &  0.10   &  \nl
  A0754&   25&  1.5 & 1.6 &  16347 &  178&   869& 156&  -0.2&  0.98&     0.75& 0.14& 0.71  &  0.19   &  \nl
  A1983&   31&  2.1 & 1.5 &  13353 &  220&  1194& 426&   0.3&  0.88&     0.72& 0.23& 1.10  &  0.63   &  \nl
  A1631&   16&  1.2 & 1.5 &  14215 &  178&   676& 139&   0.0&  1.22&     0.70& 0.14& 0.62  &  0.10   &  \nl
 Fornax&   22&  0.6 & 0.8 &   1548 &   82&   371&  52&  -0.9&  1.04&     0.84& 0.13& 0.56  &  0.07   &  \nl \nl
\tablebreak
{\bf Lenticulars} \nl  \nl
  A1656&  111&  1.9 & 0.8 &   6864 &  106&  1120&  85&   0.4&  0.46&     0.76& 0.05& 0.58  &  0.06   &  \nl
  Virgo&   81&  1.2 & 1.0 &   1418 &   76&   680&  48&  -0.6&  0.54&     0.65& 0.05& 0.46  &  0.05   &  \nl
  A3526&   85&  1.5 & 1.1 &   3501 &   96&   878&  77&  -0.6&  0.53&     0.72& 0.07& 0.65  &  0.04   &  \nl
  A0194&   35&  0.6 & 0.7 &   5397 &   57&   330&  89&  -0.8&  0.83&     0.53& 0.10& 0.54  &  0.13   &  \nl
  A0548&   52&  1.4 & 1.3 &  12415 &  110&   784&  57&  -1.0&  0.68&     0.75& 0.07& 0.50  &  0.03   &  \nl
  A1060&   49&  1.1 & 0.6 &   3795 &   93&   642&  52&  -0.8&  0.70&     0.79& 0.08& 0.60  &  0.05   &  \nl
  A2151&   36&  1.3 & 1.2 &  10855 &  130&   768&  64&  -1.2&  0.82&     0.89& 0.08& 0.56  &  0.07   &  \nl
  A1644&   47&  1.5 & 1.5 &  14207 &  124&   840& 104&   0.5&  0.71&     0.64& 0.08& 0.48  &  0.09   &  \nl
  A0539&   31&  1.6 & 0.5 &   8726 &  167&   911& 126&  -0.9&  0.88&     0.94& 0.15& 0.82  &  0.09   &  \nl
  A463s&   39&  1.0 & 1.2 &  12238 &   96&   592&  57&  -0.5&  0.78&     0.78& 0.09& 0.52  &  0.05   &  \nl
  A0496&   32&  1.3 & 0.5 &   9706 &  132&   733& 108&  -0.1&  0.87&     0.85& 0.14& 0.80  &  0.12   &  \nl
  A3376&   40&  1.4 & 1.2 &  13846 &  128&   801& 104&   0.2&  0.77&     0.79& 0.11& 0.71  &  0.10   &  \nl
  A805s&   29&  0.6 & 0.6 &   4351 &   70&   368&  39&   2.2&  0.91&     0.69& 0.09& 0.46  &  0.06   &  \nl
  A0119&   36&  1.3 & 1.1 &  13243 &  123&   729& 121&   0.5&  0.82&     0.55& 0.08& 0.56  &  0.11   &  \nl
  A0754&   25&  1.2 & 1.5 &  16072 &  140&   680&  82&  -0.9&  0.98&     0.67& 0.11& 0.50  &  0.06   &  \nl
  A1983&   27&  0.7 & 1.7 &  13681 &   83&   422& 286&   1.0&  0.94&     0.37& 0.08& 0.35  &  0.13   &  \nl
  A1631&   43&  1.3 & 1.4 &  13941 &  112&   728&  71&  -0.6&  0.75&     0.82& 0.09& 0.60  &  0.06   &  \nl
 Fornax&   24&  0.6 & 0.6 &   1507 &   68&   323&  49&  -0.2&  1.00&     0.68& 0.10& 0.51  &  0.10   &  \nl \nl
{\bf Ellipticals}  \nl  \nl
  A1656&  158&  1.6 & 0.8 &   6958 &   72&   908&  63&   1.2&  0.39&     0.61& 0.04& 0.55  &  0.05   &  \nl
  Virgo&   56&  1.2 & 0.8 &   1172 &   94&   701&  78&   0.2&  0.65&     0.61& 0.08& 0.50  &  0.06   &  \nl
  A3526&   47&  1.6 & 0.7 &   3472 &  133&   902&  66&  -1.2&  0.71&     0.86& 0.08& 0.54  &  0.04   &  \nl
  A0194&   35&  0.7 & 0.9 &   5344 &   68&   398&  66&   1.5&  0.83&     0.64& 0.11& 0.65  &  0.12   &  \nl
  A0548&   31&  1.5 & 1.3 &  12391 &  154&   841&  93&  -0.7&  0.88&     0.78& 0.08& 0.45  &  0.07   &  \nl
  A1060&   13&  1.0 & 0.6 &   3660 &  175&   596& 166&   0.0&  1.36&     0.48& 0.16& 0.49  &  0.40   &  \nl
  A2151&   12&  0.6 & 1.6 &  10522 &  114&   369&  53&  -1.4&  1.41&     0.44& 0.08& 0.47  &  0.06   &  \nl
  A1644&   19&  1.8 & 1.9 &  14252 &  254&  1067& 244&  -0.3&  1.12&     0.70& 0.18& 0.75  &  0.25   &  \nl
  A0539&   11&  1.0 & 0.2 &   8729 &  194&   601& 279&   0.4&  1.48&     0.38& 0.19& 0.56  &  0.37   &  \nl
  A463s&   22&  1.1 & 1.0 &  12495 &  143&   649& 209&  -1.3&  1.04&     0.75& 0.13& 0.32  &  0.05   &  \nl
  A0496&   27&  0.8 & 0.2 &   9818 &   88&   445& 126&   1.5&  0.94&     0.43& 0.09& 0.46  &  0.09   &  \nl
  A3376&   13&  0.7 & 0.5 &  14050 &  123&   417& 133&   0.4&  1.36&     0.33& 0.05& 0.39  &  0.22   &  \nl
  A805s&    9&  0.9 & 0.3 &   4111 &  186&   507& 151&  -0.4&  1.63&     0.70& 0.19& 0.81  &  0.30   &  \nl
  A0119&   24&  1.5 & 0.6 &  13394 &  180&   859& 349&  -0.1&  1.00&     0.42& 0.14& 0.79  &  0.42   &  \nl
  A0754&   22&  1.5 & 0.8 &  16451 &  195&   885& 167&  -0.2&  1.04&     0.77& 0.16& 0.76  &  0.24   &  \nl
  A1983&   14&  0.9 & 0.6 &  13575 &  145&   514& 111&  -0.9&  1.31&     0.57& 0.12& 0.42  &  0.11   &  \nl
  A1631&    9&  0.9 & 0.9 &  13706 &  181&   493&  85&  -1.4&  1.63&     0.60& 0.13& 0.46  &  0.14   &  \nl
 Fornax&   20&  0.5 & 0.5 &   1396 &   64&   277&  80&  -0.1&  1.10&     0.49& 0.14& 0.58  &  0.60   &  \nl
\enddata


\tablenotetext{a}{Colless and Dunn (1996)}

\tablenotetext{b}{Biviano et al.(1996)}

\tablenotetext{c}{Malumuth et al. (1992)}

\tablenotetext{d}{Fabricant et al. (1993)}

\tablenotetext{e}{Ferguson (1989)}

\end{deluxetable}

 There  were 322 objects not classified  in
the cited  references, in these cases  one of us  (AR) have classified
them  using the images from the  Digitized  Sky Survey from the STScI.
In all   cases, these galaxies are   members  of clusters  having some
objects already classified, allowing us  to compare our classification
scheme  with the  original published   classification.  Clusters  with
radial velocity but with no classification at all, as A85 or DC0107-46
were not   included  in  our analysis  because   a  good morphological
classification  was not  guaranteed.   We have  adopted  a very simple
classification   scheme   dividing  objects  in  ellipticals,    dwarf
ellipticals, spirals,  dwarf  spirals,  lenticulars,  irregulars   and
unknowns, in the case  we couldn't  find a  suitable class.  A  random
sample    of objects  was  selected  for   testing the  classification
procedure,  and  the comparison  of   our classification with  the RC3
catalog  and Dressler (1980b) is  within the  range of agreement among
traditional morphologists, 75\%  to 80\% (Andreon, 1996).  As expected
we noted  that the miss-identification  of the morphological  class is
larger in the case of lenticular galaxies.

A  mean synthetic  cluster,  (MSC),  was built by adding  the  velocity,
position and morphology information  of all galaxies from the clusters
in our sample.  The velocities  of each object  were corrected for the
mean velocity of the host cluster  and normalized by the corresponding
velocity dispersion. Their relative positions  inside the cluster were
also  normalized  using  a fiducial  radius.   Hopefully  this average
synthetic cluster preserves  the   radial dependence   of  kinematical
properties  of the sample, clearing  off the effects due to eventually
present   local  substructures.  

In  Figure  3  we   present the space
distribution,  the average deviation as   a function of the normalized
radius, and the   line-of-sight  velocity histograms of all   galaxies
separated by morphological populations of the MSC cluster. 
  The left panels from top to bottom present the projected position of
spirals, lenticulars and ellipticals. 
As  we  mention before,  the positions were   scaled using as fiducial
 radius  the so called virial radius,   $r_{200}$, used by Carlberg et
 al.  (1997) and defined as, 
$$r_{200} = \frac{\sqrt{3} \sigma}{10 H(z)} $$ 

\noindent where $\sigma$  is the global   velocity dispersion of  each
cluster  and the  Hubble  constant $H(z)$  in our case  was adopted as
H$_o$=100 kms$^{-1}$Mpc$^{-1}$.   This radius, defined as  the radius
where the mean interior  density is 200 times  the critical density of
the   universe,  is expected to  contain  the  bulk of  the virialized
cluster  mass.  The  middle panels  show,   the absolute value  of all
velocities normalized  to the  corresponding global cluster dispersion
as a function  of the normalized radius.  Using this data we estimated
the average deviation ($|u|$) within  radial rings having 100 galaxies
each (solid lines).  The dashed line  shows the expected value for the
case of isotropic orbits ($|u| \sim 0.8$).   The right panels show the
histograms of the  relative velocities normalized to the corresponding
velocity dispersion. 

  From Figure   3 we conclude that   spirals tend to present  a broader
 distribution in space and  also in  velocities distribution.  On  the
 contrary the   distribution of ellipticals  is more  concentrated, in
 both  positions  and  velocities,  reflecting   the presence  of  the
 morphological segregation inside  each individual cluster.   Although 
 there is a large spread in the velocity distribution, we observe that
 the average deviation of ellipticals as a function of radius seems to
 have values  always  lower than  the expectation  value for isotropic
 orbits indicating more eccentric orbits and showing again a different
 behavior   from  spirals. Finally, it  is    interesting to note that
 lenticular galaxies show an intermediate behavior between spirals and
 ellipticals.   We can also  observe that spirals  in the outer region
 ($r  > r_{200}$) tend to present  more eccentric orbits, as indicated
 by the   lower  values    of   the average  deviation    of  velocity
 distribution. 

To test the  normality of  the velocity  distributions, showed in  the
right    panels of Figure 3,   we  have applied the Kolmogorov-Smirnov
statistic, and  an improved W-test  (Yahil and  Vidal, 1977).  In both
cases the probability  to reject  the hypothesis is  of the  order  of
25\%.  Therefore, although the significance level is low,  we cannot
reject  that  these distributions   could  be
described by a Gaussian distribution. 

\begin{figure}[ht] 
\plotone{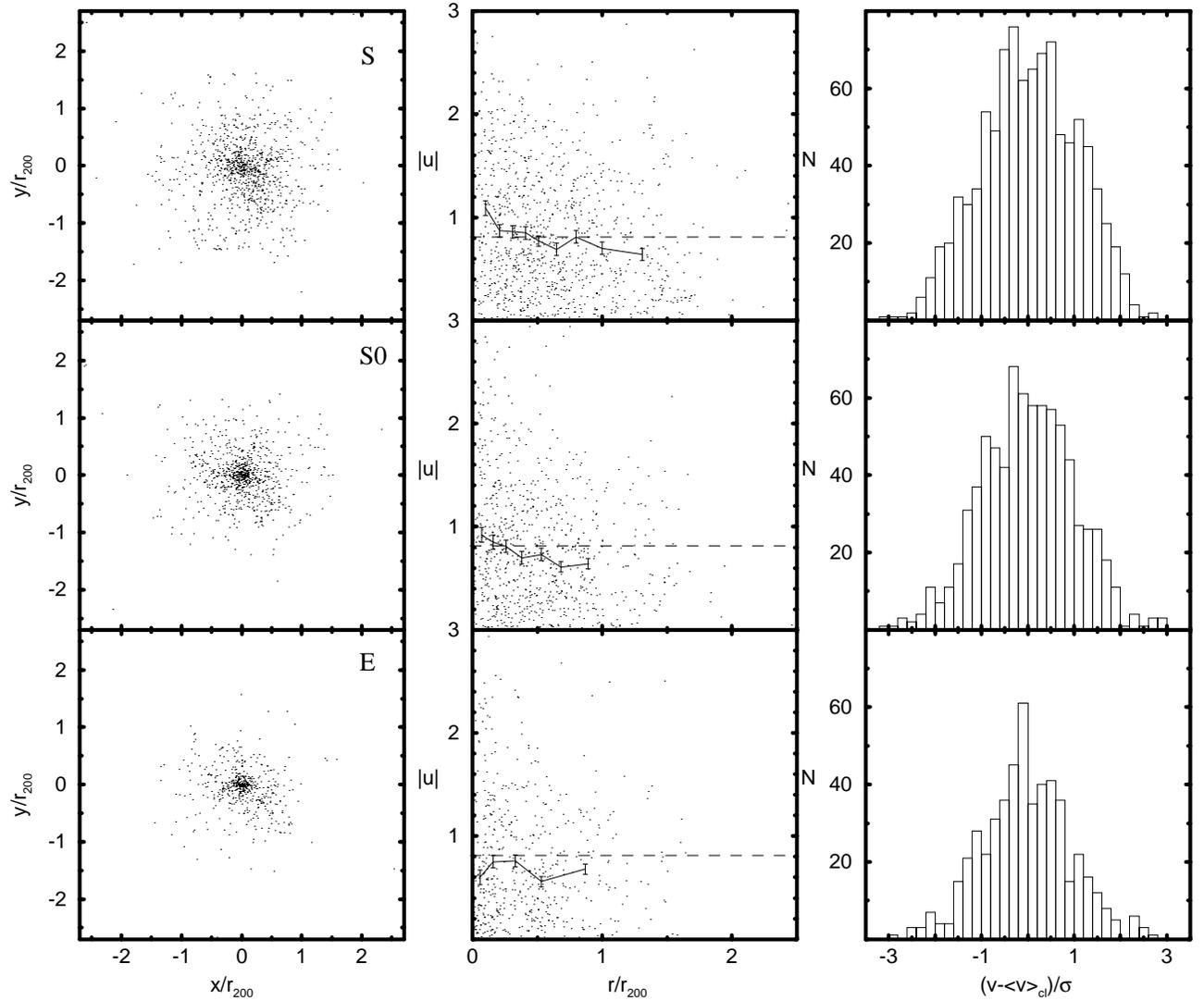} 
\caption
{ Mean Synthetic cluster (MSC) for spiral (top), lenticular (middle) and elliptical (bottom). In the 
left panels the  projected positions are in units of $r_{200}$. In the 
middle panels it is presented the modulus of the velocity normalized 
to the velocity dispersion.  The solid lines correspond to the average 
deviation within rings of 100 galaxies each, and the error bars are at the 68\%
confidence level.  Right panels:  histograms of the velocities
relative to the mean cluster velocity normalized to the projected velocity
dispersion of the corresponding cluster, bins are 200
km/s width. \label{f3}}
\end{figure}

\placefigure{f3}

The  individual properties of the clusters   are summarized in Table 1,
separated by morphological classes.  Columns (1) and (2) show the name
of the cluster  and the number of  galaxies within 2.5 $h^{-1}$Mpc. In
column (3) we present  the $r_{200}$ radius  in $h^{-1}$Mpc and column
(4) shows the harmonic radius, $r_h$,  which gives a estimation of the
mean projected  separation between  galaxies inside  each  cluster. In
columns (5)  and  (6) we show  the mean  velocity (\vbi)  and velocity
dispersion (\sbi)  in  \kms, together  with  their errors at  the 68\%
confidence   level. Columns  (7) and  (8)   show the kurtosis and  the
average  deviation  of  the line-of-sight velocity,   $|u|_{cl}$, with
their respective errors at   the  68\% confidence level.   Column  (9)
shows  the dispersion of   the cluster average deviation,  and finally
column (10) shows  the velocity references and the
other names as  the cluster is known.  Each  selected cluster has four
entries  in  this table,  one corresponding to  the  data of  all objects
irrespective of their  morphological type and  one entry  only for the
spirals, lenticulars and ellipticals. 

We have mentioned at the beginning of this section  that in some cases
a larger number  of  published velocities  are available, but  without
morphological  classification.   To   estimate  the  influence  of the
velocity incompleteness  on our mean  values, presented in Table 1, we
have also estimated the kinematical parameters using all the published
velocities for two clusters (second  lines in A496 and A805s entries).
For these two clusters the total number of galaxies is almost a factor
two  larger  than     the  number of   objects    having morphological
information.  Nevertheless, the  differences in the  average deviation
are    well   within  the error   bars,  showing     that the velocity
incompleteness does not affect drastically our  results.  In section 4
we will return to this point using the data  for Virgo and Coma, where
we have a larger number of data. 

In section 2, we show that the expected value of the average deviation
is independent of the  expected value of  the velocity dispersion.  It
means that  two populations having   the same velocity  dispersion may
present different anisotropies,  and consequently different values of
the average deviation. Therefore   we would expect that the   measured
velocity  dispersion on  Table    1  should  be independent   of   the
morphological class, if they were  subjected to the same gravitational
potential and   having a common   Gaussian  velocity distribution with
spherical symmetry.  However we notice  that there are some small, but
systematic, differences  between the velocity  dispersion of each 
morphological classes. A  possible explanation of these differences
is discussed in the section 5.  However, for our present purposes the
existence  of this difference poses a  conceptual problem since we
need   to  define the  average  deviation  normalized  by the velocity
dispersion.   In the present  analysis  we have  used  all the average
deviations  normalized to the    velocity dispersion deduced from  the
whole  cluster population.  Actually, we  repeat  the same analysis but
normalizing to the velocity dispersion of each morphological class 
and it makes no difference in our conclusions, except  for  an small
increase in  the errors. The reason, is that this  morphological
segregation   on  the  velocities dispersions is small enough to
not significantly affect the determination of the average deviation.

\section{Average deviation of the line-of-sight velocity}

In  Figure 4 we show  the histograms of  the average  deviation of the
line-of-sight velocity for the 18 clusters presented  in Table 1.  The
upper     panel represents  the  spiral   population,   the middle panel 
the
lenticulars, and the lower one the ellipticals.  In the left set of panels
we present the histograms of the   central region, around 1.0 \Mpc, of  each
cluster, while  in  the right panels the  whole 2.5  \Mpc region.  The  open
histogram  represent the contribution  of clusters were there might be
some suspicious of  low level substructures as  detected by Girardi et
al.  (1996).   The vertical dashed lines show  the  expected value for
the  extreme cases  of  radial (R)   and circular (C)  orbits for  our
Gaussian model discussed in the previous sections. 

\begin{figure}[ht] 
\plotone{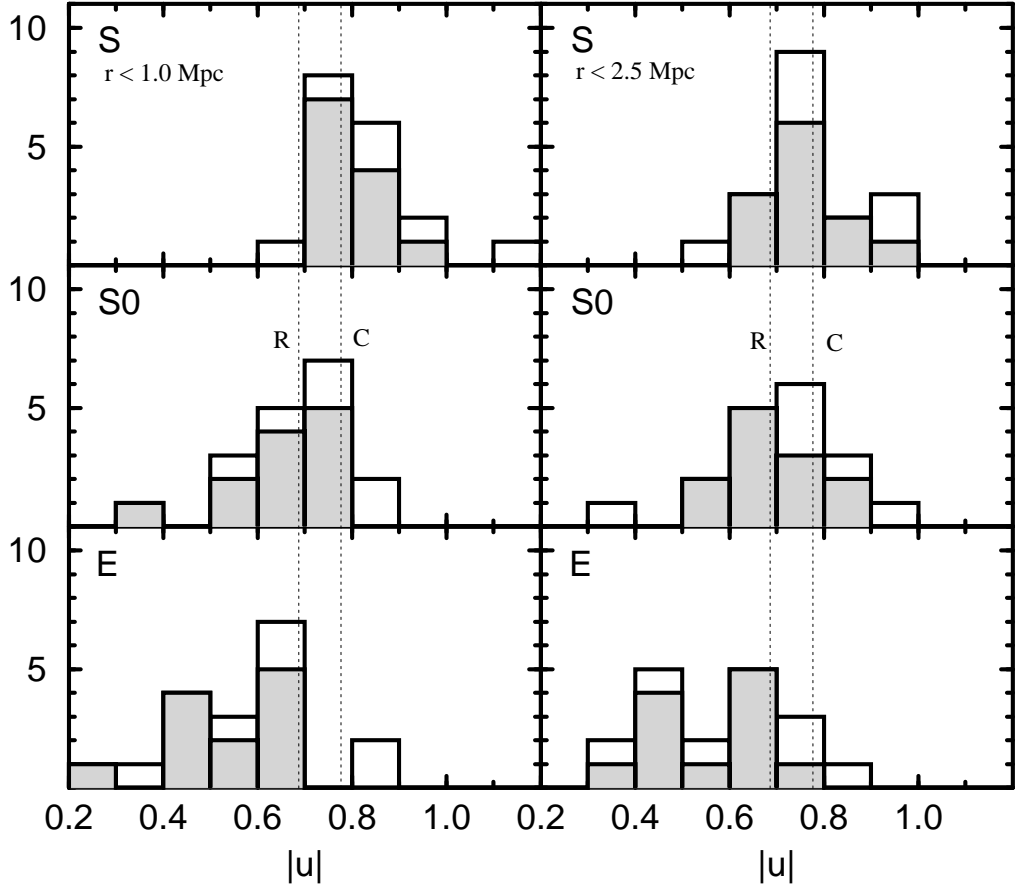} 
\caption
{The cluster sample distribution of the average deviation of the
line-of-sight velocity.  Histograms of the
spiral, lenticular and elliptical populations inside 1.0 \Mpc and 2.5 \Mpc are
plotted at right and left side, respectively.  The filled histograms represent
clusters without significant substructures, the open histograms show the
contribution of clusters were there might be some suspicious substructures as
detected by Girardi et al.  (1996).  The vertical dashed lines show the expected
value for the extreme cases of radial (R) and circular (C) orbits. \label{f4}}
\end{figure}

\placefigure{f4}

Although   all histograms   present a large    intersection region, it
remains true that  spirals tend to  peak around $\bar{|u|}  = 0.74 \pm
0.12$ in the 2.5 \Mpc  sample and $\bar{|u|} =  0.80 \pm 0.17$ in  the
1.0 \Mpc region.  These values  are close to the  prediction of a pure
circular ($|u|_{cir} \sim  0.77$), or isotropic 
($|u|_{iso} \sim  0.80$) models, although as remarked in section 2
we cannot distinguish between 
the circular case ($\eta = 0$), the isotropic case ($\eta = 1$), 
and an slightly radial case  having $\eta \sim 2$.   On the other hand
the elliptical distribution peaks in the region  $\bar{|u|} = 0.59 \pm
0.13$  and  $\bar{|u|} =  0.59 \pm 0.14$  for  2.5 \Mpc and  1.0 \Mpc,
respectively, both values close  to the prediction of the radial
orbit limit  ($|u|_{rad}  \sim  0.69$). 
We remark that this limit is the lowest value permitted by the simple Gaussian
distribution adopted in the present analysis. We can observe from Figure 4
that several clusters have mean deviations below that limit. 
This is a limitation of the present model, as will be discussed in section 5.

The difference   between the behavior   of spirals and ellipticals  
is  persistent in both,  central (1.0 \Mpc) region and the  whole 
(2.5 \Mpc) cluster.
To  further clarify this point and quantify the degree of 
significance of these differences we  have
applied some statistical   tests to verify  (1) if  the distributions
comes  from a normal  distribution  and (2)  that  distributions of
different morphological class comes from the same parent distribution.
To test the  normality we used  the modified   Kolmogorov-Smirnov
statistic, and an improved W-test (Yahil and Vidal, 1977). 
In Table 2 we  present the mean values and the  results of these two 
tests applied to the observed distribution of the average 
deviation values. Here we have in column (1) the morphological
 population, in column (2)
the  characteristic  of the   sample  considering only those  clusters
  {\em Without
substructures} and {\em All clusters}.  
In column (3)  we have the number of clusters
included in the respective sample and in column (4) the outer limiting
radius in $h^{-1}$Mpc. Again, we divided the data into one sample well inside the expected virialized region and a complementary region containing all the cluster members. Column  (5) shows the mean of the
average deviation value with the error  at the 68\% confidence level,
including the estimated observational  errors.   Column (6) shows  the
dispersion of the mean  value with  the error  at the 68\%  confidence
level.  Columns   (7) and (8)  show the   probability to   accept  the
hypothesis  of   normality, with values  expressed   in   percentage. 
The normality  tests show that   the    hypothesis  of normal
distribution of $|u|$ among clusters values cannot be rejected.

\begin{deluxetable}{llccccrr} 
\small
\tablewidth{43pc}
\tablenum{2}
\tablecaption{Average deviation in rich clusters}
\tablehead{
\colhead{Population} &
\colhead{sample}     &
\colhead{N}          &
\colhead{r$_{outer}$}&
\multicolumn{1}{c}{$<|u|>$}          &
\multicolumn{1}{c}{$\sigma_{<|u|>}$} & 
\colhead{K-S}   &
\colhead{W}     \\
\colhead{}    &
\colhead{}    &
\colhead{}    &
\colhead{Mpc} &
\multicolumn{1}{c}{}    &
\multicolumn{1}{c}{}    &
\colhead{\%}  &
\colhead{\%}  
}
\startdata
{\bf S+S0+E}& All clusters          & 18 & 2.5 & 0.73 ~~0.08 & 0.07  (-0.01,+0.02) & 25 & 31 \nl
            &                       & 18 & 1.0 & 0.73 ~~0.08 & 0.07  (-0.01,+0.01) &  5 & 50 \nl
            &                       &    &     &              &                     &    &    \nl
            & Without substructures & 12 & 2.5 & 0.71 ~~0.07 & 0.05  (-0.02,+0.01) & 25 & 81 \nl
            &                       & 12 & 1.0 & 0.71 ~~0.07 & 0.06  (-0.01,+0.02) &  5 &  7 \nl
            &                       &    &     &              &                     &    &    \nl
{\bf S}     & All clusters          & 18 & 2.5 & 0.74 ~~0.12 & 0.09  (-0.02,+0.01) &  5 & 5  \nl
            &                       & 18 & 1.0 & 0.80 ~~0.17 & 0.10  (-0.01,+0.03) & 25 &15  \nl
            &                       &    &     &              &                     &    &    \nl
            & Without substructures & 12 & 2.5 & 0.75 ~~0.14 & 0.07  (-0.02,+0.02) & 10 & 8  \nl
            &                       & 12 & 1.0 & 0.81 ~~0.17 & 0.08  (-0.01,+0.02) & 25 &33  \nl
            &                       &    &     &              &                     &    &    \nl
{\bf S0}    & All clusters          & 18 & 2.5 & 0.73 ~~0.09 & 0.13  (-0.02,+0.03) & 25 &61  \nl
            &                       & 18 & 1.0 & 0.72 ~~0.13 & 0.12  (-0.02,+0.03) & 25 & 9  \nl
            &                       &    &     &              &                     &    &    \nl
            & Without substructures & 12 & 2.5 & 0.70 ~~0.11 & 0.11  (-0.03,+0.02) & 25 &60  \nl
            &                       & 12 & 1.0 & 0.69 ~~0.11 & 0.11  (-0.02,+0.03) & 25 & 4  \nl
            &                       &    &     &              &                     &    &    \nl
{\bf E}     & All clusters          & 18 & 2.5 & 0.59 ~~0.13 & 0.16  (-0.02,+0.04) & 25 &69  \nl
            &                       & 18 & 1.0 & 0.59 ~~0.14 & 0.16  (-0.02,+0.04) & 25 &65  \nl
            &                       &    &     &              &                     &    &    \nl
            & Without substructures & 12 & 2.5 & 0.57 ~~0.15 & 0.14  (-0.04,+0.03) & 25 &73  \nl
            &                       & 12 & 1.0 & 0.55 ~~0.15 & 0.13  (-0.02,+0.04) & 10 &17  \nl
\enddata

\end{deluxetable}

From Figure 4 and the average values of mean deviation presented
 in Table 2, we
conclude that   ellipticals have  a  mean deviation  indicating more
eccentric orbits than spirals. This result is of high significance and
is valid both in the inner and also in the whole cluster inside the
 2.5 \Mpc region.
 A key question  consists, despite of the difference in
 $|u|$  in    Table 2,  in  asking   if   the distributions   of  each
 morphological class are originated from the same parent distribution. 
 In particular we are interested to see if the distribution of 
 elliptical galaxies deviations could be drawn from a population of
 spiral galaxies deviations. To
 test this hypothesis we have applied the two-sampling tests developed
 in IRAF/STSDAS ({\em  twosampt} and {\em  kolmov}), using the unbinned 
 data.   The results are
 summarized  in  Table   3,  where column (1)    shows the  two tested
 populations. Column (2)  and (3) show the  samples defined  using the
 same criteria as  in Table 2.  Column  (4) shows the outer  radius in
 $h^{-1}$Mpc, column  (5)  shows the result of  the Kolmogorov-Smirnov
 test, the results are  expressed  as  the  probability that  the  two
 compared population came from  the same parent distribution.  Column
 (6) shows the result of the Gehan's Generalized Wilcoxon test,
 again the results are expressed as the probability of both population
 came  from   the    same   distribution,   values   are   given    in
 percentage. Column (7)  presents  the results of Log-rank  test
 and column (8) the results of the Peto \& Peto Generalized
 Wilcoxon test.  The above statistical tests allow us to conclude that
 $\bar{|u|}$ values   of  ellipticals  and  spirals  indeed  come   from two
 different parent distributions, while for spirals and lenticulars they
 came from different population only inside the inner 1.0 \Mpc region.

\begin{deluxetable}{llllllll} 

\small
\tablewidth{43pc}
\tablenum{3}
\tablecaption{Two-sampling test between populations}
\tablehead{
\colhead{Populations} &
\colhead{sample} &
\colhead{N} &
\colhead{r$_{outer}$} &
\colhead{KS} &
\colhead{GGW} &
\colhead{L}  &
\colhead{PPGW} \\
\colhead{}       &
\colhead{}      &
\colhead{}      &
\colhead{Mpc}      &
\colhead{\%} &
\colhead{\%} &
\colhead{\%} &
\colhead{\%} 
}
\startdata

{\bf S x E} & All clusters          & 18 & 2.5  & 0.2   &  0.1  & 0.02 &  0.7    \nl
            &                       & 18 & 1.0  & 0.001 &  0    & 0    &  0.0001 \nl
            &                       &    &      &       &       &      &         \nl
            & Without substructures & 12 & 2.5  & 0.2   &  0.05 & 0    &  0.03   \nl
            &                       & 12 & 1.0  & 0.001 &  0.0  & 0.00 &  0.01   \nl
            &                       &    &      &       &       &      &         \nl
{\bf S x S0}& All clusters          & 18 & 2.5  & 27    &  39   & 20   &  76     \nl
            &                       & 18 & 1.0  & 6     &  0.4  & 0.2  &   2     \nl
            &                       &    &      &       &       &      &         \nl
            & Without substructures & 12 & 2.5  & 10    &  16   &  3   &  46     \nl
            &                       & 12 & 1.0  & 3     &  0.2  & 0.1  &   1     \nl
            &                       &    &      &       &       &      &         \nl
{\bf E x S0} & All clusters          & 18 & 2.5  & 6     &  2    & 3    &  6      \nl
            &                       & 18 & 1.0  & 6     &  2    & 2    &  6      \nl
            &                       &    &      &       &       &      &         \nl
            & Without substructures & 12 & 2.5  & 10    &  2    & 1    &  8      \nl
            &                       & 12 & 1.0  & 10    &  2    & 7    &  10     \nl
\enddata

\end{deluxetable}
   
In Coma (402 objects) and  Virgo (458 objects)  we could apply a finer
analysis,   since  these  two   clusters have  the   largest number of
velocities and morphological data, in our sample.  Moreover, they also
deserve  an special attention since they  are representative of nearby
clusters  rich in ellipticals  and spirals, respectively.   In a first
step we  have divided  the data  for these two   clusters into  a {\em
``core"} and a   {\em ``halo"} regions, in    order to test   eventual
differences in these two environments.  The separation radius dividing
these two regions  was determined  by the  condition of both  having a
similar number  of  objects, and  not by  fitting the density  profile
itself.  So  that the  {\em ``core"}  represents  the  densest central
region, while the {\em ``halo"}  the  outer low  density regime.   The
result of this analysis is   presented in Table    4, divided by 
morphological  family.  We  observe  that in  both regions remains the
trend of  ellipticals having lower $\bar{|u|}$  than spirals.  In both
clusters we observe that $  \bar{|u|}_E \simeq 0.61$, remarkably close
to the limit of radial orbit, found in our model.  In contrast spirals
have $\bar{|u|}_S \simeq 0.80$, closer   to the circular or   slightly
radial case.

\begin{deluxetable}{llrr@{}rr@{}rrr}  

\tablewidth{40pc}
\small
\tablenum{4}
\tablecaption{Kinematic parameters of Coma and Virgo clusters}
\tablehead{
\colhead{Sample: covered radius} &
\colhead{population} &
\colhead{N}       &
\multicolumn{2}{c}{v$_{BI}$}       &
\multicolumn{2}{c}{$\sigma _{BI}$ }&
\colhead{ K}            &
\colhead{$|u|$}         \nl
\colhead{}    &
\colhead{}    &
\colhead{}    &
\multicolumn{2}{c}{km/s}    & 
\multicolumn{2}{c}{km/s}    &
\colhead{}    &
\colhead{}              }
\startdata
{\bf Coma}     \nl
{\em all members}:0.0 - 2.0 [Mpc] &  S+S0+E&   458&   6969&  48&  1032&  39&   0.8 ~~~0.2&     0.71  ~~~0.03\nl
                                  &  S     &   189&   7036&  79&  1082&  66&   0.7 ~~~0.4&     0.73  ~~~0.05\nl
                                  &  S0    &   111&   6864& 106&  1120&  85&   0.4 ~~~0.5&     0.76  ~~~0.05\nl
                                  &  E     &   158&   6958&  72&   908&  63&   1.2 ~~~0.4&     0.61  ~~~0.04\nl 
\nl
{\em core}:0.0 - 0.5 [Mpc]        &  S+S0+E&   247&   6918&  73&  1153&  59&   0.3 ~~~0.3&     0.70  ~~~0.04\nl
                                  &  S     &    94&   7024& 121&  1172&  98&  -0.1 ~~~0.5&     0.72  ~~~0.06\nl
                                  &  S0    &    63&   6877& 165&  1305& 133&   0.1 ~~~0.6&     0.76  ~~~0.08\nl
                                  &  E     &    90&   6854& 109&  1030&  96&   0.9 ~~~0.5&     0.61  ~~~0.06\nl
\nl
{\em halo}:0.5 - 2.0 [Mpc]        &  S+S0+E&   211&   7036&  61&   894&  50&   1.6 ~~~0.3&     0.72  ~~~0.04\nl
                                  &  S     &    95&   7060& 103&   997&  85&   1.8 ~~~0.5&     0.78  ~~~0.07\nl
                                  &  S0    &    48&   6863& 131&   896&  83&  -0.6 ~~~0.7&     0.80  ~~~0.05\nl
                                  &  E     &    68&   7084&  91&   748&  73&   0.2 ~~~0.6&     0.59  ~~~0.06\nl
\nl
{\bf Virgo}    \nl
{\em all members}:0.0 - 2.5 [Mpc] &  S+S0+E&   402&   1380&  39&   785&  23&  -0.6 ~~~0.2&     0.76  ~~~0.03\nl
                                  &  S     &   265&   1414&  51&   824&  26&  -0.8 ~~~0.3&     0.82  ~~~0.03\nl
                                  &  S0    &    81&   1418&  76&   680&  48&  -0.6 ~~~0.5&     0.65  ~~~0.05\nl
                                  &  E     &    56&   1172&  94&   701&  78&   0.2 ~~~0.7&     0.61  ~~~0.08\nl
\nl
{\em core}:0.0 - 0.8 [Mpc]        &  S+S0+E&   213&   1149&  53&   774&  31&  -0.7 ~~~0.3&     0.76  ~~~0.04\nl
                                  &  S     &   122&   1135&  76&   841&  44&  -1.1 ~~~0.4&     0.89  ~~~0.05\nl
                                  &  S0    &    47&   1229&  91&   620&  72&   0.0 ~~~0.7&     0.55  ~~~0.07\nl
                                  &  E     &    44&   1082& 108&   706&  89&   0.0 ~~~0.7&     0.63  ~~~0.09\nl
\nl
{\em halo}:0.8 - 2.5 [Mpc]        &  S+S0+E&   189&   1629&  52&   710&  24&  -1.2 ~~~0.4&     0.84  ~~~0.03\nl
                                  &  S     &   143&   1630&  61&   728&  29&  -1.1 ~~~0.4&     0.85  ~~~0.04\nl
                                  &  S0    &    34&   1676& 118&   674&  49&  -1.4 ~~~0.8&     0.82  ~~~0.08\nl
                                  &  E     &    12&   1467& 194&   629& 135&  -0.6 ~~~1.4&     0.62  ~~~0.12\nl
\enddata

\end{deluxetable}

\normalsize

In Figure 5 we show the radial behavior of the kurtosis,  velocity 
dispersion and the average deviation for  these two  clusters and
also for the MSC cluster. The continuous line refers to spirals, while the dashed line represents the ellipticals. In the Virgo cluster we used rings containing 20 object, while in Coma they contain 25 object. In the MSC as stated
 before the rings were defined to contain 100 objects.
The velocity  dispersion in the MSC cluster was  scaled to 1000 \kms ~to
ease the comparison with the corresponding plots for Virgo and Coma. 
We observe that the  velocity dispersion has the same behavior and hence
basically contains  the   same information as the average deviation. However, 
 by using  the average  deviation we could link the 
orbital model and the line-of-sight velocity distribution while the same
 is not possible when we use the velocity dispersion.
In  addition, the velocity  dispersion error is larger than the
 average deviation error, resulting therefore in 
 a higher statistical significance.
We further observe that 
the velocity dispersion and the  average deviation profiles of MSC are 
in good agreement with the corresponding profiles of Coma and Virgo. 
Moreover, we can also observe that the velocity dispersion is
 a decreasing function of the radius, in the outer
regions of all three samples.

 We remark that the kurtosis profile is  very noisy, even in the
 MSC cluster making it rather difficult to extract some useful 
 information from  this diagram.

\begin{figure}[ht] 
\plotone{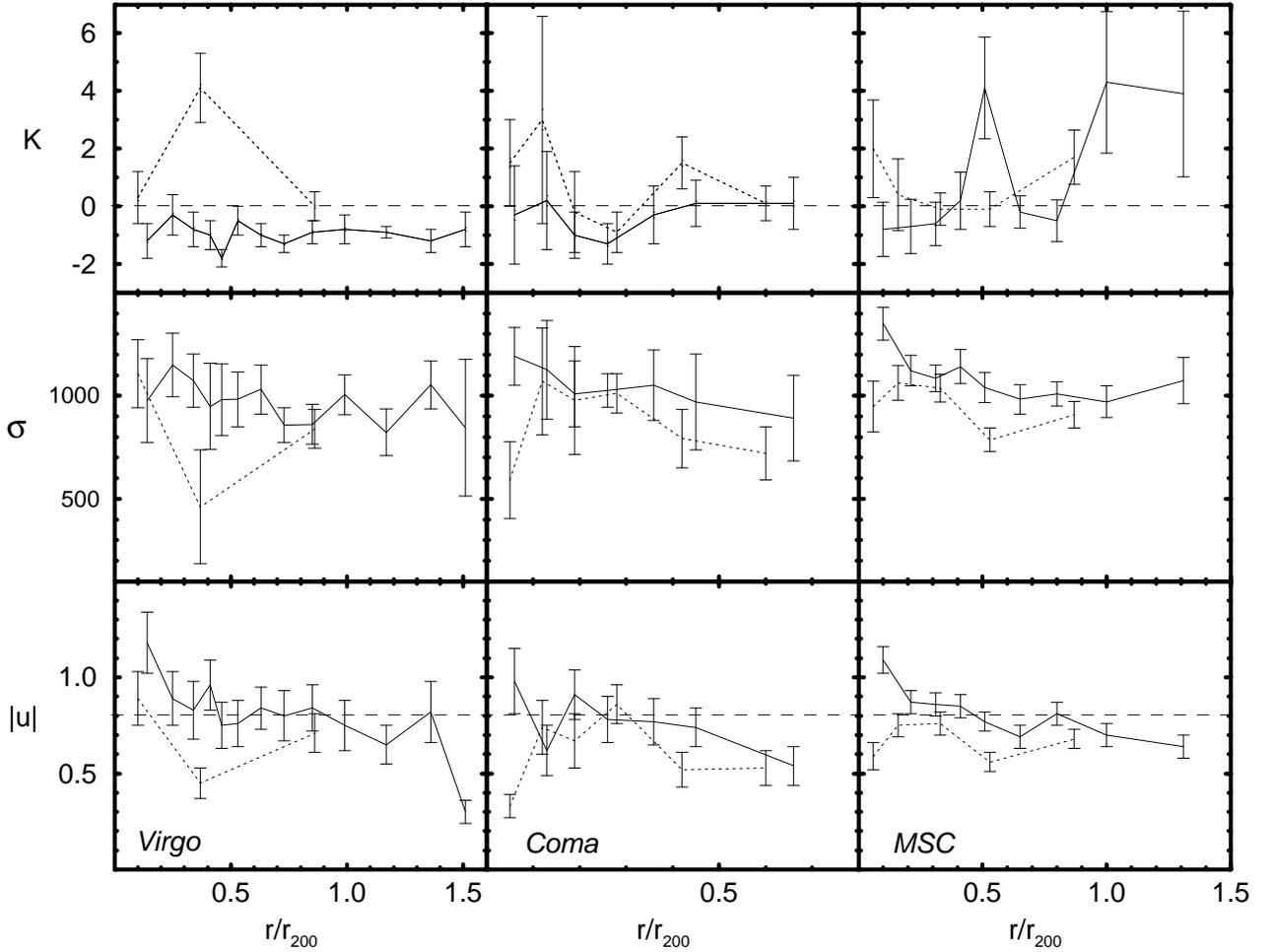} 
\caption
{The kurtosis, velocity dispersion and average deviation  
  as a function of the radius for Virgo,  Coma and the MSC clusters.  
  The radius is  expressed in unit of $r_{200}$.
 Estimatives of these parameters  were done using rings with 20, 25 and 100 galaxies, in Virgo, Coma  and MSC respectively.
 Solid  lines represent the spirals and dotted 
lines the ellipticals.  Error bars are at the 68\%
confidence level. The long dashed line presented  in the kurtosis and average deviation plots correspond to the expectation value  when orbits 
are isotropic  \label{f5}}
\end{figure}

\placefigure{f5}

 Therefore, the tendency of ellipticals having lower $|u|$ is found in
all the samples. In fact, this effect is present both 
in the 1.0 \Mpc samples and also in the sample at 2.5  \Mpc. The
clusters  with substructures and  without
substructures show the same effect and also the individual clusters
as Virgo and Coma. We remark that the  same effect  is found in the MSC  
 as is shown in Figures 3  and 5. 

 Therefore we conclude
that ellipticals  tend to show  a systematically lower value of $|u|$,
implying in more eccentric orbits, in contrast  with spirals that 
present higher  values, related with a more circular or isotropic
orbit.  Clusters with substructures  tend to  follow the same  trend,
although with a larger  dispersion.  Moreover, it is  interesting to
observe  that lenticular  galaxies   tend to present  an  intermediate
behavior between ellipticals and spirals.  We remark that these trends
are   also present,  even   if we normalize  the   distribution to the
velocity   dispersion of each morphological     family instead of  the
velocity dispersion of all the cluster. 

It is interesting compare our conclusions with  the work of Biviano et
al.  (1992). They have compiled a sample  of 68 clusters   with at least 30
galaxies separating the behavior  of galaxies more luminous than
the third ranked galaxy ($m_3$). They found that the brighter galaxies
are preferentially located in  the  central regions having in average   a
morphological type of $<T> = -3$, while the less luminous ones have an
average type $<T> = -1$.  They notice that this effect is not induced by
morphological    segregation,   and   is    not     restricted  to  cD
clusters.  Moreover   it   does not   depend    on  the   presence  of
substructures. They propose that the energy equipartition status seems
to be    achieved  by these   low velocity   galaxies.   Following our
analysis, their Figure 2 shows a $|u|$ value of 0.55 $\pm$ 0.05 to the
galaxies brighter than the third ranked galaxy  and the other galaxies
have a $|u|$ of 0.81 $\pm$ 0.01. Then, a value near to circular orbits
fit very well  the $m >  m_3$ galaxies, while for the brighter one 
we need  more eccentric orbits.

\section{Velocity dispersion profiles}

As already pointed
out by other authors, in most clusters the velocity dispersion of
spirals  tend to   be larger than  the  one  observed in  ellipticals.
Actually, 13  out   of the 18  clusters   of the present   sample show
this effect.

\begin{figure}[ht] 
\plotone{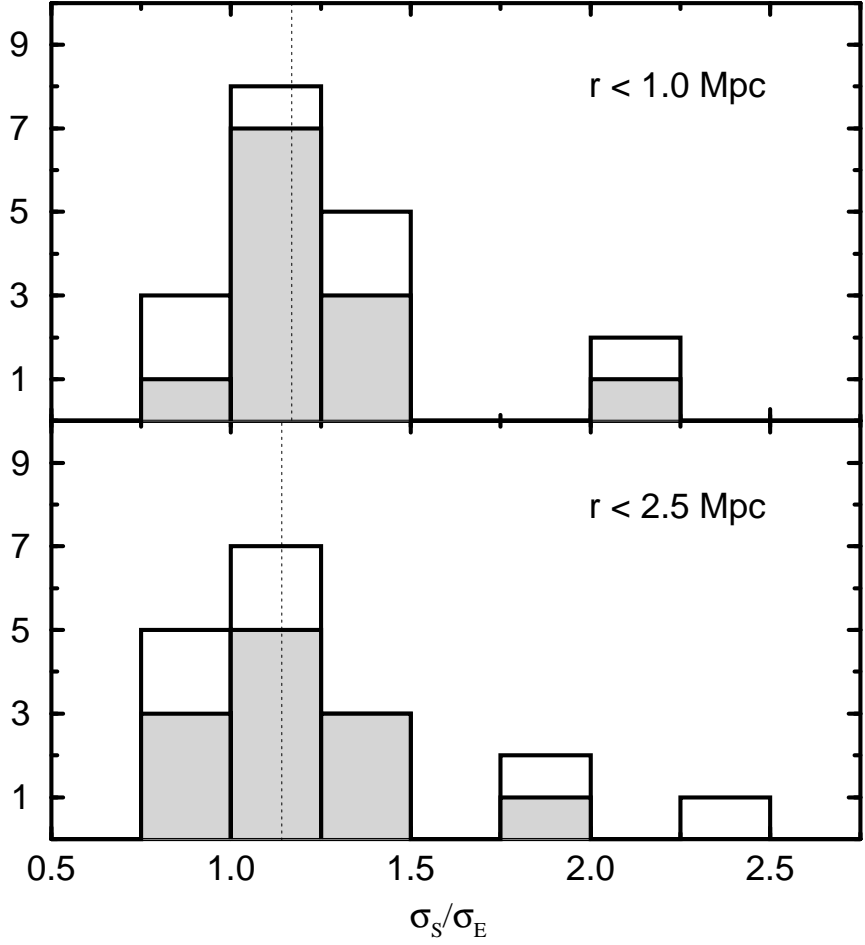} 
\caption
{Histogram of the ratio of velocity dispersions between spirals and ellipticals in 18
nearby clusters.  Upper  plots show the distribution considering all
members inside 1.0 \Mpc, while the data for 2.5 \Mpc is in the lower panels.
   The two vertical lines
correspond to the median values. \label{f6}}
\end{figure}

\placefigure{f6}

In Figure  6 we  present an histogram  of  the  ratio of  the velocity
dispersions  of both populations,  $\sigma_S/\sigma_E$, sampled at two
different radii.  We can  observe that in  the internal ($<$ 1.0 \Mpc)
and  external regions  ($<$ 2.5 \Mpc),  there  is  a tendency  for the
velocity dispersion of spirals being larger.  The two dotted vertical
lines   correspond  to the  median  value  computed  on  basis  of the
individual cluster  values. Again the  filled  histogram correspond to
the cluster  without   substructures.    For  the  1.0   \Mpc   region
$<\sigma_S/\sigma_E> \simeq 1.17 \pm 0.06$ with  a dispersion of $0.25
\pm  0.12$, while for 2.5   \Mpc  region we have  $<\sigma_S/\sigma_E>
\simeq 1.14 \pm 0.08$ and a dispersion of $0.31 \pm 0.11$.  Therefore,
the average velocity dispersion  of spirals is $\sim$15\%  larger than
the corresponding velocity dispersion of  ellipticals, but with a high
 variance. 

From the analysis presented on section 3 we predicted that 
if both systems
respond to the same gravitational potential their 
velocity dispersion  would have the same value,  even if
the anisotropy parameter could be  different. Therefore, these 
observed differences in  velocity 
dispersion could imply that the initial supposition of a Gaussian 
distribution simply could be not true, or that both populations are not feeling the same gravitational potential.
Another possibility is that this effect  in  the   velocity  
dispersion could  arise    from
differences in the dynamical  state, virialized or falling models.
Moreover, there might
also be   some contribution  due  to  the presence  of substructures
 inside each population.  

The  existence  of differences  in the dynamical
state due to the possibility that spirals  are  recently falling members 
 captured by  the
system and ellipticals correspond to a collapsed and virialized
system, would imply in a ratio  of  1.41 between  the
 velocity   dispersion of both populations.
  This  would
happen because,  for spirals, the minimum   energy to be bound  could be
expressed as $E_S = 0$, resulting that $T_S=-W_S$.   On the other hand
if ellipticals  are virialized  then $2T_E=-W_E$.  Therefore,  we obtain 
that $ W_S/W_E = T_S/2T_E  $, while if both  systems are responding to  the
same potential we conclude that $ T_S  = 2T_E$.  If both families
are affected in  the same  way by projection  effects, having  the same
kind of  orbits  and the same  spatial symmetry,  we would expect 
that
$\sigma_S = 1.41  \sigma_E$.  In fact,  some authors  (Colless \&  Dunn,  1996,
Andreon, 1996) have used these argument favoring the idea that spirals
are falling into  the cluster. However,  we remark from our analysis 
that   spirals have  a   more
isotropic rather than radial orbits.  
Considering the  orbit information from our
analysis  it  remains  possible that spirals  are  in  the  process of
capture, but in  that  case they are  not ``falling"  in radial orbits
into the cluster core, except perhaps for the outer ones, as can be seen in Figure 3.

An alternative explanation for the difference between the velocity dispersion of spirals and ellipticals 
 is that  they  are spatially 
segregated inside     the  cluster,   they could be submitted to  different
gravitational potentials.  If we consider the density profiles ($\rho =
\rho_o /  r(r+b)^2 $) recently discussed   by Navarro et  al.  (1995),
then we could  predict that two segregated families,  at  radii $R_E$ and
$R_S$, should present virialized velocity dispersion in the ratio, 

\[ \frac{\sigma_S}{\sigma_E} = \left( \frac{1 -\frac{ln(R_S/b +
1)}{R_S/b}}{1 - \frac{ln(R_E/b + 1)}{R_E/b}} \right) ^{1/2} \]

\noindent where  $b$ stands for the core  radius of the cluster (0.3 -
0.5 \Mpc).   In  Coma for example  we would conclude, using  the sample
of objects  with
radial velocities,  that  the  harmonic radius  among spirals
would be 1.01 \Mpc, while ellipticals have 0.83 $h^{-1}$Mpc.  Therefore , on
basis of these estimatives we would predict  $\sigma_S /\sigma_E = 1.04
- 1.05$, where the different values stands for  differences in the core
radius.   Since for  Coma $\sigma_S  /\sigma_E =   1.19  \pm 0.09$, we
conclude that quite  probably the  observed difference is not affected
by this effect. 

Projection effects  and sub clustering  could  also have some influence on
the observed velocity distributions.  In fact, Cen (1997) using N-body
simulation in   a CDM universe  model   obtains  a  more  quantitative
estimatives of these effects.  His results suggest that the presence of
substructures modifies the velocity distribution in a complex way, but
the  final  velocity   dispersion  is slightly    affected. He estimated
variations of the final velocity dispersion  of only 5\%, 9\% and 27\%
within 0.5,  1.0 and 2.0  \Mpc, respectively.  Another study   using
observational data is presented in Bird et al. (1996). They noted that
the existence of substructures is an important factor to determine the
dynamical  parameters, but the  effect  is reduced  when the data  is
restricted to a region inside the  virial radius. In  our case, we are
working with  most of the galaxies  well inside  of the virial radius.
This   supposition is  respalding  by  the  fact that $r_{200}$  in all
clusters present  values always higher than  the  harmonic radius. 
  Finally, it is worthwhile mention that variations
on the velocity dispersion due to substructures could increase or reduce
the velocity dispersions (Bird,  1994), a  fact which is   not consistent with
spirals always presenting an slightly larger   dispersion  
 than   ellipticals.  

Finally, we investigate if the difference in the velocity dispersion is due
to the  possibility that ellipticals  and   spiral  follow a
different velocity distribution than a Gaussian.
To analyze this point we referred the
works of Kent \&  Gunn (1982), The \& White  (1986), Malumuth  \& Kriss
(1986), Merritt \& Saha (1993).  They use different velocity and density
distributions assuming different potentials, and they study 
the  behavior  of  the    velocity  dispersion  with the 
radius, $\sigma(r)$, instead of the total  dispersion. 

Kent and Gunn (1982)  found that the best  fit description for a rich cluster
as Coma is achieved by
using either an isotropic King-Michie distribution, or a constant anisotropic
 model.  In this later case when a higher degree of anisotropy is  allowed  
the  models, for the same cluster, tend to present a  lower velocity
dispersions.  This trend is in agreement with our findings that
 ellipticals have more eccentric orbits and lower velocity dispersion
 than spirals.  In  fact, from Figure  5 we can observe that in Virgo, 
 Coma  and also in the MSC  we have
$\sigma_E(r)  < \sigma_S(r)$. 

An additional constraint to the velocity distribution comes from the probable
existence of a dark matter component that dominates the mass of the
cluster. In this respect 
Merritt  and  Saha   (1993) have investigated an interesting
 method to recover the
gravitational potential of a cluster making use of measured  line-of-sight
velocity data. An application was done for the Coma cluster assuming 
that the potential is dominated by a dark matter halo. The first solution
 corresponds to a mass
distribution which is slightly more concentrated than the galaxies themselves. In that case the velocity distribution is nearly  radial 
inside $\sim$  1.0 \Mpc  region, and isotropic  outside. On the other
hand two other solutions were analyze corresponding to a singular
isothermal sphere, with a high density core, and an smooth halo density. 
In that case the orbital distribution is isotropic inside  0.5 \Mpc and 
circular  outside. 
None of these three  models
allow for the presence of radial orbits at large
radius, mainly because this solution would required a lower velocity 
dispersion.  An alternative model to explain a velocity  dispersion 
 exceeding 600 \kms 
 ~in the
outer parts of Coma, was previously investigated by The \& White (1986)
that   found equilibrium models 
fitting  the data,  assuming it as  being  
dominated by a massive
core   of $\sim 10^{15} M\odot$. Using these   models  they  were able  to fit the
velocity dispersion data assuming  that galaxies  in  the outer part of  the
cluster  have  nearly  circular     orbits.

 It is  worthwhile mention  that although a power-law model is not a good description for Coma (Kent and Gunn, 1982), nevertheless  there are clusters where this
model gives a good description to the data. An example is the poor cluster MKW4
 analyzed by Malumuth and Kriss (1986). In this particular case the steep 
dropping of the velocity dispersion profile is very well reproduced by the anisotropic power-law model. We remark however that in the present study most of our sample of rich clusters do not show such behavior and hence the constant anisotropic model, or isotropic King-Michie model, is more satisfactory.

In summary, since  our sample is dominated by relatively rich clusters, the
observed differences of the velocity dispersion among spirals and
ellipticals is consistent with the models discussed by Kent and Gunn (1982).
In this respect we favor the scenario where ellipticals have more anisotropic orbit distribution and lower dispersion. One the contrary, spirals are more
 representative of a population having a more isotropic, or even circular, 
orbit distribution and larger velocity dispersion. On possible example that
illustrate these differences is the distribution of the velocity modulus
for a Plummer potential shown by Merritt and Saha (1993). In fact their
Monte-Carlo simulation of radial orbit models is similar distributed as the
out data for ellipticals in the MSC, while their circular
model have the same trend as the one we observe in spirals.

\section{Conclusions}

From the previous  analysis we conclude that elliptical galaxies in rich regular
clusters have more eccentric orbits than spirals.  This effect implies
that ellipticals are passing more  often in the dense central  regions
suffering therefore  a larger influence   due to tidal effects.  In  a
first order approximation  we  can  imagine  a  cluster modeled by   a
central mass concentration  and surrounded by test particles.
If we consider a typical particle  having  $v_R \simeq \sigma_R$ and
$v_\perp  \simeq \sqrt{2}\sigma_\perp$.   Energy  and angular momentum
conservation would require that the orbit  of this object should be  bound
to  the region   $R/R_G = 1   \pm \frac{\eta}{\sqrt{2+\eta^2}}$, where
$R_G$  is the  cluster  gravitational radius.   For an isotropic orbit
($\eta = 1$) this typical object will be oscillating between $0.42 \le
R/R_G \le  1.58$.  However in the case  of a typical  elliptical, with
$\eta  \rightarrow   \infty  $, we have   $0.0   \le R/R_G \le  2.00$.
Therefore our  average  elliptical will be  more  exposed to the tidal
interacting field of the central core objects. 

The  preference of ellipticals  for radial   orbits could  be  a
 plausible explanation why we often find two bright ellipticals in the
 central region of clusters.  These pairs of ellipticals of comparable
 masses  are called dumbbell galaxies   (Matthews, 1964).  Examples of
 this type of objects are  NG4782/3 or the  central pair in A3266  (de
 Souza  \& Quintana  1990,  Quintana et  al. 1996a,b). Almost  25\% of   the
 brightest cluster galaxies are multiple systems, and from a sample of
 116 types BM I-II clusters extracted from the Abell catalog 51 of them 
 have dumbbell as BCM
 (Gregorini et  al.  1994).   The  ultimate  evolution of a   dumbbell
 system will be  the formation a  larger central object, probably a  cD
 galaxy (Tremaine,  1990).  It   is  interesting to observe  that  the
 presence of an spiral forming a pair in the  central region of a rich
 cluster is a  very rare event.  This  could be a probable consequence
 of   the  different orbital  parameters  of  spiral and elliptical in
 clusters. 

There are basically  two scenarios that  could explain the kinematical
segregation  found in the present  work.  We can imagine that clusters
are  generated  from large amplitude   perturbations.  After  the core
formation objects   in the outskirts   of  the cluster, preferentially
spirals, are gradually accreted forming  therefore an halo of  objects
having more circular orbits than  the original objects first collapsed
in the core most of them ellipticals.  On the other  extreme we may as
well imagine  a situation were  large virialized groups collides given
rise to   a  knew cluster, larger than the   former  one, as the  dynamical
behavior of A2151 (Bird, 1995), or Coma (Biviano et  al.  1996) in the
BCG formation   scenario presented by  West (1994).    In that case  we
expect that galaxies with low angular momentum will  tend to cross the
central regions and possibly may  experience large tidal interactions
changing their morphologies.

\acknowledgments

We thank  to the referee, E. Malumuth,  for valuable comments. Also we
gratefully acknowledge financial support  from CAPES (AR fellowship)
and  FAPESP (RdS grant No 1995/7008-76). This  research has made use of
data obtained through the  High  Energy Astrophysics  Science  Archive
Research Center  Online Service,  provided  by the  NASA/Goddard Space
Flight Center.

\newpage
\onecolumn

\end{document}